\documentclass[useAMS,usenatbib]{mn2e}

\usepackage{graphicx,subfigure}
\usepackage{amsmath}
\usepackage{txfonts}

\title[Gamma-Ray Bursts: the Isotropic-Equivalent-Energy Function and the Cosmic Formation Rate]{Gamma-Ray Bursts: the Isotropic-Equivalent-Energy Function and the Cosmic Formation Rate}
\author[\emph{S.-W. Wu, D. Xu, F.-W. Zhang and D.-M. Wei}]{Shi-Wei Wu$^{1,2}$\thanks{swwu@pmo.ac.cn},
Dong Xu$^{3}$\thanks{dong.xu@weizmann.ac.il}
Fu-Wen Zhang$^{1,2,4}$
and Da-Ming Wei$^{1,2}$\thanks{dmwei@pmo.ac.cn}
\\
$^{1}$Purple Mountain Observatory, Chinese Academy of Sciences, Nanjing 210008, China\\
$^{2}$Key Laboratory of Dark Matter and Space Astronomy, Chinese Academy of Sciences, Nanjing 210008, China\\
$^{3}$Benoziyo Center for Astrophysics, Weizmann Institute of Science, Rehovot 76100, Israel\\
$^{4}$College of Science, Guilin University of Technology, Guilin 541004, China}
\begin{document}

\date{Accepted 2012 April 05. Received 2012 April 04; in original form 2012 February 04}

\pagerange{\pageref{firstpage}--\pageref{lastpage}} \pubyear{2002}

\maketitle

\label{firstpage}

\begin{abstract}
Gamma-ray bursts (GRBs) are brief but intense emission of soft $\gamma-$rays, mostly lasting from a few seconds to a few thousand seconds. For such kind of high energy transients, their isotropic-equivalent-energy ($E_{\rm iso}$) function may be more scientifically meaningful when compared with GRB isotropic-equivalent-luminosity function ($L_{\rm iso}$), as the traditional luminosity function refers to steady emission much longer than a few thousand seconds. In this work we for the first time construct the isotropic-equivalent-energy function for a sample of 95 bursts with measured redshifts ($z$) and find an excess of high-$z$ GRBs. Assuming that the excess is caused by a GRB luminosity function evolution in a power-law form, we find a cosmic evolution of $E_{\rm iso}\propto(1+z)^{1.80^{+0.36}_{-0.63}}$, which is comparable to that between $L_{\rm iso}$ and $z$, i.e., $L_{\rm iso}\propto(1+z)^{2.30^{+0.56}_{-0.51}}$ (both $1\sigma$). The evolution-removed isotropic-equivalent-energy function can be reasonably fitted by a broken power-law, in which the dim and bright segments are $\psi(E_{\rm iso})\propto E_{\rm iso}^{-0.27\pm0.01}$ and $\psi(E_{\rm iso})\propto E_{\rm iso}^{-0.87\pm0.07}$, respectively ($1\sigma$). For the cosmic GRB formation rate, it increases quickly in the region of $0 \leq z \lesssim 1$, and roughly keeps constant for $1\lesssim z \lesssim 4$, and finally falls with a power index of $-3.80\pm2.16$ for $z\gtrsim 4$, in good agreement with the observed cosmic star formation rate so far.
\end{abstract}


\section{Introduction}
Gamma-ray bursts (GRBs) are among the brightest cosmological explosions in the universe, mostly lasting from a few seconds to a few thousand seconds in soft $\gamma$-ray. Thanks to quick follow-up observations in optical band, the redshifts of some GRBs have been measured by detecting the absorbtion lines of their afterglows or the emission lines of their host galaxies. So far the number of {\it Swift}-detected GRBs with known redshifts has grown up to about one hundred and thus makes a reliable statistical analysis possible. Among various statistical works, the luminosity function as well as the cosmic formation rate of GRBs are particularly interesting. The luminosity function is a measure of the number of bursts per unit luminosity, which sheds light on the energy release and emission mechanism of GRBs. The cosmic formation rate is a measure of the number of events per comoving volume and time, which can help us understand the production of GRBs in various stages of the universe.

The isotropic-equivalent luminosity ($L_{\rm iso}$) function of GRBs was firstly assumed to be a standard candle \citep{b26,b31} and later more realistic shapes of the luminosity function were derived \citep{b32,b33,b34,b39,b40}. The cosmic GRB formation rate has also been extensively investigated \citep[e.g.,][]{b4,b5,b6,b7,b8,b9,b32,b15,b36}. For example, \citet{b4} found a correlation between the variability degree of the prompt gamma-ray light curve and the luminosity and then adopted it to estimate the luminosities/redshifts of 220 bright long GRBs detected by {\it CGRO}/BATSE. \citet{b7} used this GRB sample to estimate the luminosity function evolution and the cosmic formation rate of GRBs. Wei \& Gao (2003) find that there is a tight correlation between the peak energy of the prompt emission spectrum ($E_{\rm peak}$) and the luminosity. Recently, \citet{b38} find that both short and long GRBs all comply with this correlation. Using this correlation, \citet{b9} estimated the luminosities/redshifts of the 689 BATSE GRBs and hence derived their luminosity function and the formation rate. Essentially these two works with simulated GRB redshifts reached quite similar results, i.e., the GRB formation rate increases quickly in the region of  $ 0 \leqslant z \leqslant 1$ and keeps constant up to $z \sim 10$, which is inconsistent with the cosmic star formation rates (SFRs) inferred from UV, optical, and infrared observational data so far \citep{b11,b12,b13,b14}.

The original concept of luminosity function comes from astrophysical objects such as stars and galaxies which are long-lasting and quite stable in releasing their energy. For GRB-like high energy transients, the total isotropic-equivalent energy ($E_{\rm iso}$) released in the whole duration of one event can be reliably measured, and its function (i.e.,  the number density of bursts per $E_{\rm iso}$ interval) likely provides an independent or even more representative clue on the underlying physics. That's why in this work we focus on the so-called ``isotropic-equivalent-energy function'' rather than the traditional luminosity function.

This paper is arranged as follows. \S~2 introduces our sample and data selection. \S~3 presents the statistical technique while \S~4 shows the results. We adopt a robust, nonparametric statistical technique to derive the isotropic-equivalent-energy function and the cosmic formation rate of GRBs from a $E_{\rm iso}-z$ sample. For comparison, the results from a $L_{\rm iso}-z$ GRB sample are also presented. In \S 5, we discuss the implication of our results and compare the cosmic GRB formation rate with the observational cosmic star formation rate. Throughout the paper, we use the standard $\Lambda$ cold dark matter cosmology with the typical parameters $\Omega_{\rm m} =0.27$, $\Omega_{\Lambda} =0.73$, and ${\rm h}=0.7$.

\section[]{Data analysis}
In this work, two sets of data are analyzed. The $E_{\rm iso}-z$ sample comes from \citet{b1,b2}, containing 95 long GRBs and X-Ray Flashes (XRF, i.e. particularly soft bursts). This sample is made up of two parts. The first part consists of 70 long GRBs from \citet{b1} and their redshifts range from 0.033 to 6.3. The $E_{\rm iso}$ values in this work are slightly different from those in \citet{b1} because of the different cosmological parameters adopted in two works. The second part is from \citet{b2} without any modification. All the GRB spectra have been extrapolated and corrected to [1,10000] keV in the cosmological restframe.

The $L_{\rm iso}-z$ sample is from \citet{b15} ($L_{\rm iso}$ represents the isotropic peak luminosity, and the time resolution is 1 $s$). Due to the {\em Swift}/BAT narrow energy band, only a small fraction of bursts have a well determined spectrum. In order to obtain reasonable estimates of $L_{\rm iso}$ for all bursts, \citet{b15} considered the characteristic Band function, i.e.,
\begin{equation}
N(E)=
\left\{
\begin{array}{l}
A(\frac{E}{100{\rm keV}})^\alpha \exp(-\frac{E}{E_0}),~~~~~~~~~~~~~~~~~~~~~~{\rm for}~E\leq (\alpha - \beta)E_0\\
A(\frac{E}{100{\rm keV}})^\beta [\frac{(\alpha - \beta)E_0}{100{\rm keV}}]^{\alpha - \beta} {\rm exp}(\beta - \alpha),~~~{\rm for}~E\geq (\alpha - \beta)E_0\\
\end{array}
\right.
\end{equation}
where the characteristic parameters $(E_{\rm peak},~\alpha,~\beta)$ taken as $(511~{\rm keV},~-1,~-2.25)$, respectively. To account for the so-called k-correction, again all spectra have been extrapolated and corrected to [1,10000] keV in the cosmological restframe. To test whether the above Band function applicable to all bursts, \citet{b15} preformed Monte-Carlo simulation in order to compare GRBs having simulated spectraal parameters with GRBs having measured spectral parameters. The result of such a simulation demonstrates robustness of the $L_{\rm iso}$ sample adopted above.

\section{Statistical techniques}
The $E_{\rm iso}-z$ sample suffers from various selection effects \citep{N8,b35}, among which the dominated one is the truncation due to the detection limit of the telescope, as seen in Fig. 1. If this bias not removed, the $E_{\rm iso}-z$ correlation would be far from the intrinsic one. A nonparametric $\tau$ statistical technique may be introduced to resolve this problem, which was first put forth by \citet{b16} and further developed by \citet{b17}. \citet{b7} first applied this technique to GRBs with simulated redshifts and later \citet{b9} applied it to a larger GRB sample still with simulated redshifts but up to $z\sim 10$. In this work, we continue to use this technique, but for the first time to two GRB samples with observed/measured redshifts. In short, this nonparametric $\tau$ statistical technique uses a well-defined truncation criterion to estimate the correlation (if any) between the relevant variables and their underlying parent distributions.

The $E_{\rm iso}$ and $z$ are not independent. Without loss of generality, the total isotropic-equivalent-energy function can be rewritten as $\Phi(E_{\rm iso},z)=\rho(z)\phi(E_{\rm iso}/g(z))/g(z)$, where $\rho(z)$ is the GRB formation rate at $z$, $\phi(E_{\rm iso}/g(z))$ is the present-day (i.e., $z=0$) isotropic-equivalent-energy function, and $g(z)$ counts for the cosmic evolution of $E_{\rm iso}$, that said, $E'_{\rm iso}=E_{\rm iso}/g(z)$. In the following analysis, the evolution $g(z)$ will be removed from the $E_{\rm iso}$ sample; after that one obtains the $E'_{\rm iso}$ distribution, then the cumulative function $\psi(E'_{\rm iso})$, and finally the GRB formation rate $\rho(z)$.

\begin{figure}
\centering
\includegraphics[width=9cm]{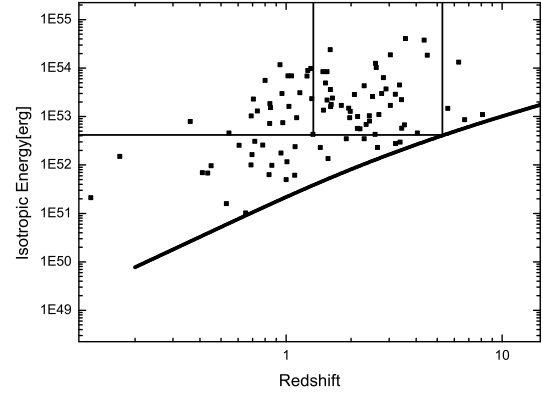}
 \caption{The $E_{\rm iso}$ distribution and the fluence limit. \label{fig1}}
\end{figure}

Consider a set of observable $E_{{\rm iso},i}$ and $z_{i}$, where $i$ indexes the $i$th burst and in our case $i$ runs from 1 to 95. As shown in Fig. 1, for the $i$th sample of ($z_{i}, E_{{\rm iso},i}$), we consider an associated set of
\begin{equation}
J_{i}=\{j|E_{{\rm iso},j}>E_{{\rm iso},i},z_{j}<z_{i,\rm lim}\},~~~~~~~{\rm for}~1\leq i\leq 95.
\end{equation}
in which the number of samples in the $J_{i}$ set is $N_{i}$. The $z_{i,\rm lim}$ is the redshift of the crossing point between two lines of $E=E_{{\rm iso},i}$ and the fluence limit corresponding to its ``isotropic-equivalent-energy" limit. If $z_{i}$ and $E_{{\rm iso},i}$ are independent to each other, one would expect the number of the following sample
\begin{equation}
R_{i}={\rm Number}\{j\in J_{i}|z_{j}\leq z_{i}\}
\end{equation}
to be uniformly distributed between 1 and $N_{i}$. To estimate the correlation degree between $E_{\rm iso}$ and $z$, one may introduce the test statistic $\tau$ parameterized as
\begin{equation}
\tau=\frac{\sum_{i}(R_{i}-E_{i})}{\sqrt{\sum_{i}V_{i}}},
\end{equation}
where $E_{i}=(N_{i}+1)/2$ and $V_{i}=(N_{i}^2-1)/12$ are the expected mean and the variance of the uniform distribution, respectively.

If $R_{i}$ follows an ideal uniform distribution, then the samples of $R_{i}\leq E_{i}$ and $R_{i}\geq E_{i}$ should be equal, and thus the statistic parameter $\tau$ tends to be zero. Note that the $\tau$ value here has been normalized by the square root of variance, so the correlation degree $z$ and $E_{\rm iso}$ can be measured in units of standard deviation.

The two variables $E_{\rm iso}$ and $z$ are connected through the form of $E_{\rm iso}=4\pi {\cal F} D^2_L(z)/(1+z)$, where ${\cal F}$ is the fluence in 1$-$10000 keV in the burst restframe. Note that bursts in the sample were actually detected by instruments onboard different satellites. The flux sensitivities of these instruments are as follows: ${\sim 4 \times 10^{-8}{\rm erg}\,{\rm cm}^{-2}\,{\rm s}^{-1}}$ for {\em CGRO}/BATSE; ${\sim 10^{-7}{\rm erg}\,{\rm cm}^{-2}\,{\rm s}^{-1}}$ for Konus-Wind, BeppoSAX and {\em Fermi}/GBM;  ${\sim 3 \times 10^{-8}{\rm erg}\, {\rm cm}^{-2}\,{\rm s}^{-1}}$ for {\em HETE-2}; and ${\sim 10^{-8}{\rm erg}\,{\rm cm}^{-2}}\,{\rm s}^{-1}$ for {\em Swift}/BAT. Some bursts were detected by more than one satellite. To be safe, we choose the best sensitivity of ${\sim  10^{-8}{\rm erg}\,{\rm cm}^{-2}\,{\rm s}^{-1}}$ as the flux limit of the whole sample.

For the $L_{\rm iso}$ sample, the two variables are connected by $L_{\rm iso}=4\pi {F} D^2_L(z)$ where ${F}$ is the flux in 1$-$10000 keV in the burst restframe. We set the flux limit as $F=5\times10^{-8}{\rm erg}\, {\rm cm}^{-2}\,{\rm s}^{-1}$ in view of the fact that the sensitivity of BAT is a few $\sim 10^{-8}{\rm erg}\,{\rm cm}^{-2}\,{\rm s}^{-1}$ \citep{b30}. Nevertheless, our results are not sensitive to the limit of ${\cal F}$ and/or $F$.

\section{Results}
Following \citet{b18} and \citet{b7}, we take the form of $g_{\rm k}(z)=(1+z)^{k}$ in order to separate the isotropic energy evolution $g_{\rm k}(z)$ from the GRB sample. The value $E'_{\rm iso}\equiv E_{\rm iso}/g_{\rm k}(z)$ represents the isotropic energy after removing the evolution effect. When $\tau$ is not equal to zero, we change the $k$ values until $\tau=0$ with a proper ${\rm k}$. Fig. 2 shows the $\tau$ value as a function of ${k}$. The null hypothesis of the evolution is rejected at about 3.5 $\sigma$ confidence level. The best fit to the $E_{\rm iso}-z$ data yields that $E'_{\rm iso}=E_{\rm iso}/(1+z)^{1.80}$, i.e., $k=1.80$.

\begin{figure}
\centering
\includegraphics[width=9cm]{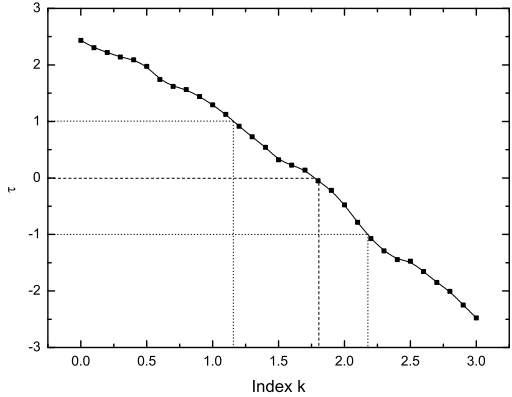}
 \caption{The statistical $\tau$ as a function of the evolution parameter ${k}$. The best-fit $\tau=0$ as well as its $1\sigma$ error (i.e., $\tau=\pm1$) correspond to ${k}=1.80^{+0.36}_{-0.63}$. Thus, $g_{k}(z)=(1+z)^{1.80}$ is the best function to describe the isotropic energy evolution. A hypothesis of no evolution (equivalent to ${k}=0$) is rejected at a significance of $2.5\sigma$. \label{fig2}}
\end{figure}

After converting $E_{\rm iso}$ into $E'_{\rm iso}=E_{\rm iso}/(1+z)^{1.80}$, we can nonparametrically derive the cumulative (local) ``isotropic- equivalent-energy function'' $\psi(E'_{{\rm iso},i})$ with the following equation of  \citep{b16,b17,b19,b7,b9}
\begin{equation}
\ln\psi(E'_{{\rm iso},i})=\sum_{j<i}{\ln(1+\frac{1}{N_{j}})}.
\end{equation}

\begin{figure}
\centering
\includegraphics[width=9cm]{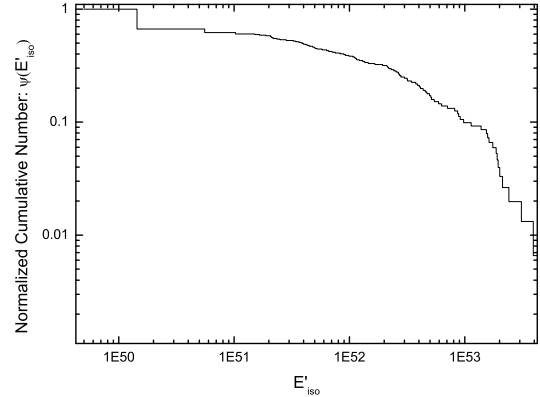}
 \caption{Cumulative isotropic-equivalent-energy function $\psi(E'_{\rm iso})$ of $E'_{\rm iso}=E_{\rm iso}/(1+z)^{1.80}$, which is normalized to unity at the dimmest point. \label{fig3}}
\end{figure}

As can bee seen, the cumulative number at the $i$th point is calculated from $N_{j}$ and for each point indexed by $j$, a truncation parallel to the axes is made and a weight $1/N_{j}$, based on the number of points in the associated set, is assigned to that data point.

Fig. 3 shows the ``isotropic-equivalent-energy function'' of $E'_{\rm iso}=E_{\rm iso}/(1+z)^{1.80}$. The shape of the ``isotropic-equivalent-energy function'' roughly follows a broken power law, and the dim and bright segments can be parameterized as $\psi(E_{\rm iso})\propto E_{\rm iso}^{-0.27\pm0.01}$ and $\psi(E_{\rm iso})\propto E_{\rm iso}^{-0.87\pm0.07}$. It corresponds to the isotropic energy distribution at $z=0$ since the evolution effect has been removed. The ``isotropic-equivalent-energy function'' in the comoving frame is $\psi(E'_{\rm iso})(1+z)^{1.80}$.

To estimate the cosmic GRB formation rate, the cumulative number distribution $\psi(z)$ as a function of $z$ is derived using the function analogous to equation (5). In this case, for the $i$th sample the associated set is given by
\begin{equation}
J'_{i}=\{{j}|z_{j}<z_{i},E_{{\rm iso},j}>E_{{\rm iso},i,{\rm lim}}\},
\end{equation}
where $E_{{\rm iso},i,{\rm lim}}$ is calculated at the crossing point of the fluence limit and $z=z_{i}$. The resulting cumulative GRB formation rate $\psi(z)$ is shown in Fig. 4.

\begin{figure}
\centering
\includegraphics[width=9cm]{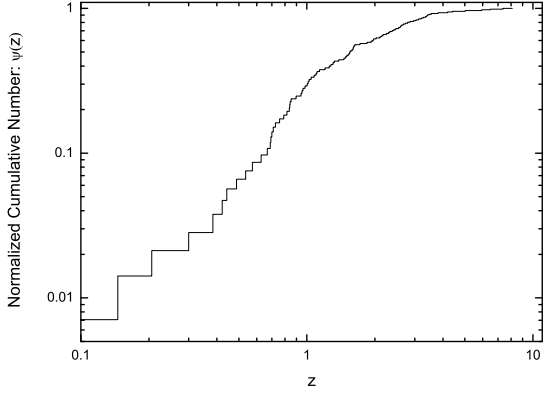}
 \caption{Cumulative GRB formation rate $\psi(z)$ as a function of $z$, which is also normalized to unity at the highest point. \label{fig4}}
\end{figure}

To be scientifically useful, one needs to convert the cumulative formation rate into the differential form (e.g., to compare with the cosmic star formation rate). The conversion is given by
\begin{equation}
\rho(z)=\frac{d\psi(z)}{dz}(1+z)[\frac{dV(z)}{dz}]^{-1},
\end{equation}
where $(1+z)$ comes from the cosmological time dilation, and $dV(z)/dz$ is a differential comoving volume described by
\begin{equation}
\frac{dV}{dz}=4\pi(\frac{c}{H_0})^3[\int^z_0\frac{dz}{\sqrt{\Omega_{\Lambda}+\Omega_{\rm m}z^3}}]^2\times\frac{1}{\sqrt{\Omega_{\Lambda}+\Omega_{\rm m}z^3}}.
\end{equation}

\begin{figure}
\centering
\includegraphics[width=9cm]{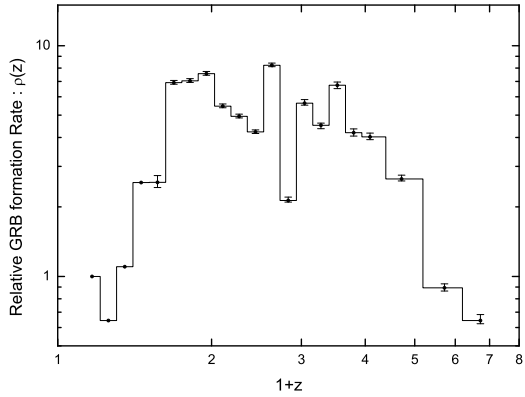}
 \caption{Relative GRB formation rate, which is normalized to unity at the first point. The error bar represents the 1 $\sigma$ statistical uncertainty of each point. \label{fig5}}
\end{figure}

Fig. 5 shows the resulting differential GRB formation rate. The best-fit power-laws for different segments are
\begin{equation}
\rho(z)\propto
\left\{
\begin{array}{l}
(1+z)^{8.24\pm4.48}~~~~~~~~~{\rm for}~z<1\\
(1+z)^{-0.54\pm0.64}~~~~~~~{\rm for}~1<z<3.5\\
(1+z)^{-3.80\pm2.16}~~~~~~~{\rm for}~z>3.5\\
\end{array}
\right.
\end{equation}
with $95\%$ confidence bounds.

The $L_{\rm iso}-z$ sample was treated in the same way. For this sample, we found ${k}=2.30^{+0.56}_{-0.51}$, which is close to the value from the $E_{\rm iso}-z$ sample. The corresponding luminosity function, cumulative GRB formation rate and differential form of the GRB formation rate have been reported in Figs. 6-8.

\begin{figure}
\centering
\includegraphics[width=9cm]{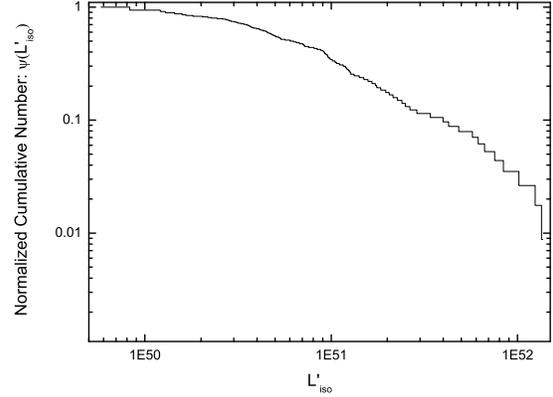}
 \caption{Cumulative luminosity function $\psi(L'_{\rm iso})$ of $L'_{\rm iso}=L_{\rm iso}/(1+z)^{2.30}$, which is normalized to unity at the dimmest point. The evolution effect is removed. \label{fig6}}
\end{figure}

\begin{figure}
\centering
\includegraphics[width=9cm]{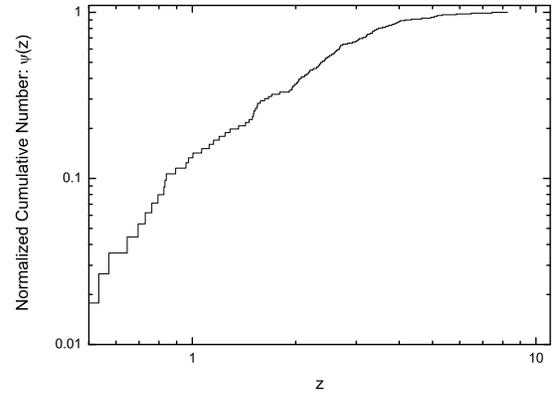}
 \caption{Cumulative GRB formation rate of $L_{\rm iso}$ sample as a function of $z$, which is normalized to unity at the highest point. \label{fig7}}
\end{figure}

\begin{figure}
\centering
\includegraphics[width=9cm]{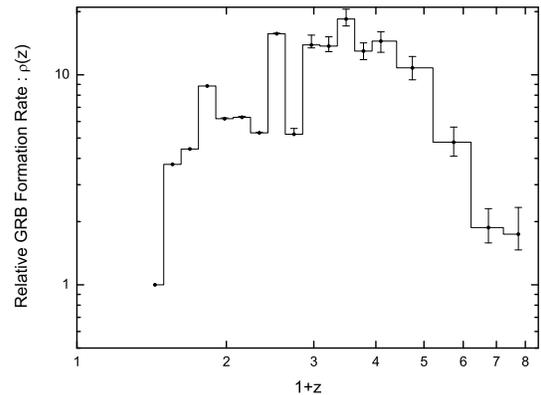}
 \caption{Relative GRB formation rate of $L_{\rm iso}$ sample, which is normalized to unity at the first point. The err bar represents the 1 $\sigma$ statistical uncertainty of each point. \label{fig8}}
\end{figure}

We use the best-fit isotropic-equivalent-energy function and GRB formation rate to calculate log N-log S distributions and compare them with the observed log N-log S distributions of BATSE and {\it Swift} bursts. The results are shown in Fig. 9. From this figure, we find that the results obtained from our model can well reproduce the observation (The probabilities of the K-S test are P=0.95 and P=0.74 for the BATSE and {\it Swift} GRB samples, respectively).

\begin{figure}
\centering
\includegraphics[width=9cm]{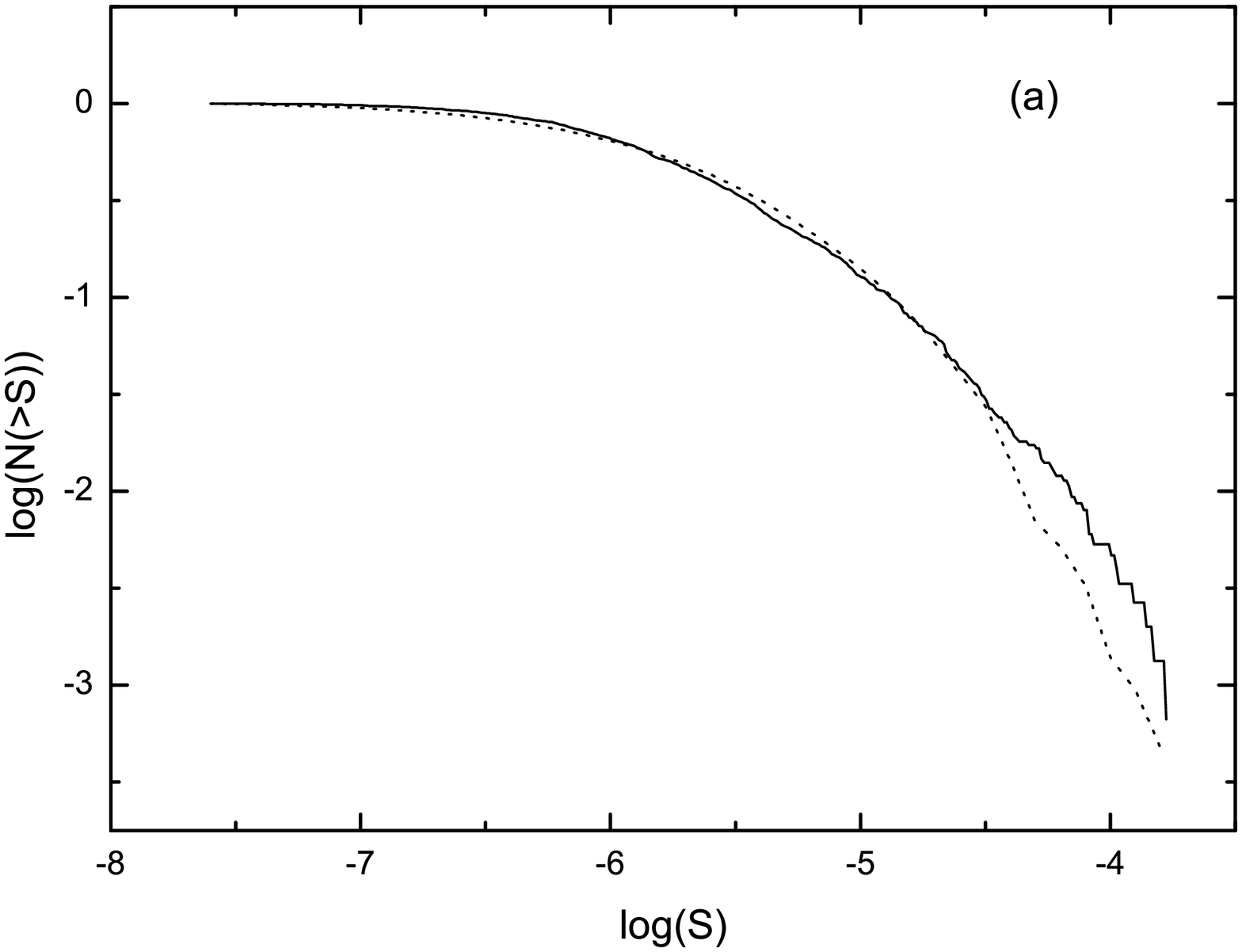}
\includegraphics[width=9cm]{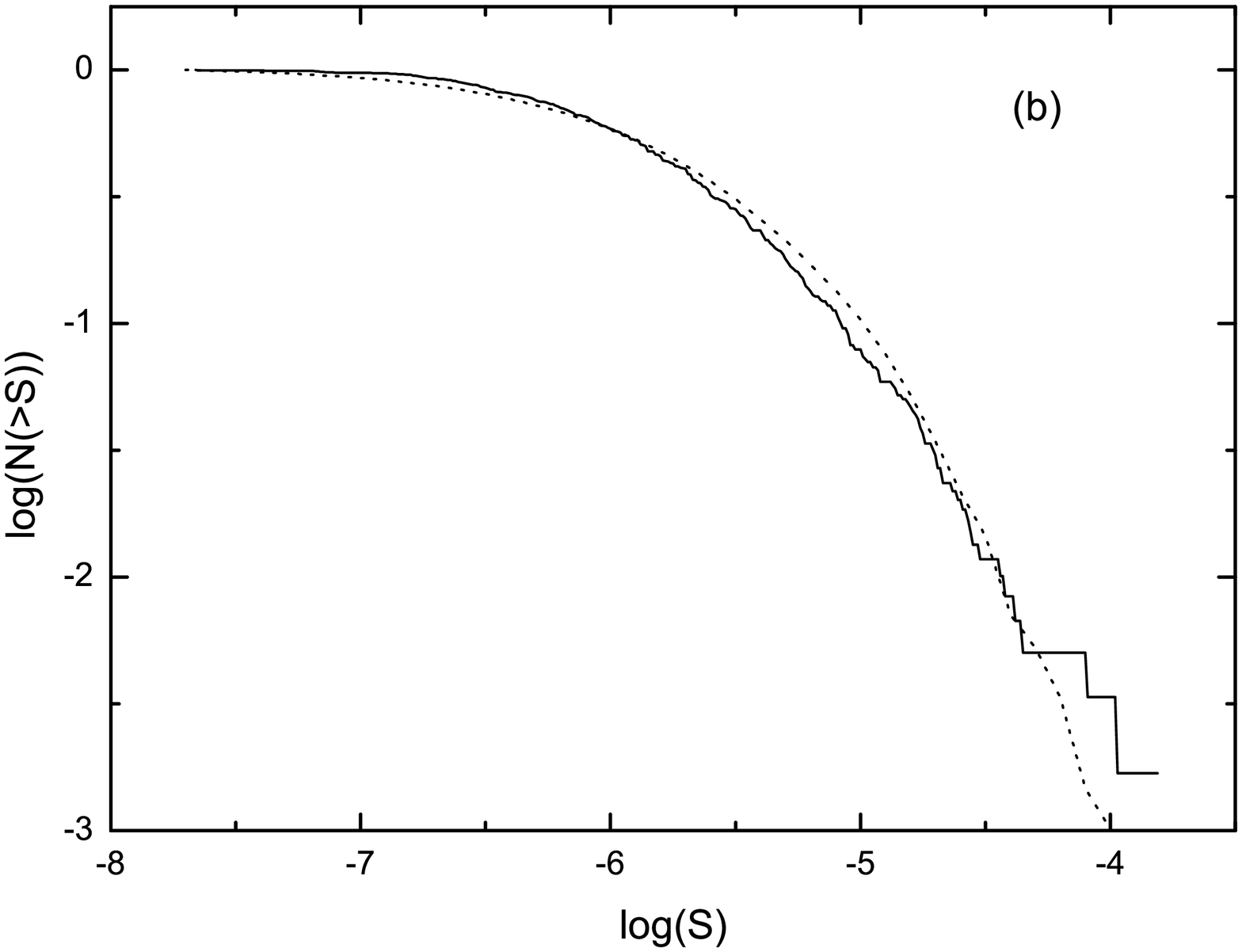}
\caption{Cumulative bursts distribution as a function of the fluence (S). Panel (a): log N-log S distribution for BATSE bursts (Solid line) and the distribution predicted by our model (Dotted line). Panel (b): log N-log S distribution for {\it Swift} bursts (Solid line) and the distribution predicted by our model (Dotted line).   \label{fig9}}
\end{figure}

\section{Conclusion and Discussion}
GRBs are brief but intense emissions of soft $\gamma-$ray, lasting from a few seconds to a few thousand seconds. For GRB-like high energy transients, the total isotropic-equivalent energy, $E_{\rm iso}$, can be reliably measured, and the number density of bursts per $E_{\rm iso}$ interval may provide an independent or even more representative clue on the underlying physics of GRBs. In this work, using a sample containing 95 bursts with measured redshifts, we for the first time constructed the isotropic-equivalent-energy function and then adopted it to estimate the GRB formation rate. The fluence-truncation effect has been properly addressed by adopting a $\tau$ statistical technique. We find there exists cosmic evolution between $E_{\rm iso}$ and $z$, i.e., $E_{\rm iso}\propto g_{\rm k}(z)=(1+z)^{1.80^{+0.36}_{-0.63}}$ (see Fig.~2), which is comparable with that  between $L_{\rm iso}$ and $z$, i.e., $L_{\rm iso}\propto(1+z)^{2.30^{+0.56}_{-0.51}}$. Our finding of the evolution is largely consistent with previous findings using GRB samples with simulated redshifts \citep{b7,b10,b9}.

We also find that the evolution power index $k$ is {\em not} sensitive to the chosen value of fluence/flux as long as it is around the instrumental sensitivity. Changing the fluence limit from $ 8\times 10^{-7}{\rm erg}\,{\rm cm}^{-2}$ to $ 5\times 10^{-8}{\rm erg}\, {\rm cm}^{-2}$, the best-fit index $k$ of $E_{\rm iso}$ sample only varies from 1.80 to 2.05, changing the flux limit from $ 5\times 10^{-8}{\rm erg}\, {\rm s^{-1}} {\rm cm}^{-2}$ to $ 1\times 10^{-8}{\rm erg}\, {\rm s^{-1}} {\rm cm}^{-2}$, the best-fit index $k$ of $L_{\rm iso}$ sample only varies from 2.30 to 2.65, all well within the 1$\sigma$ uncertainty.

As can been seen, the evolution of $E_{\rm iso}$ is comparable with that of $L_{\rm iso}$. This might indicate that the durations of GRBs do not evolve significantly with redshift and that the existence of the isotropic-equivalent-energy evolution may reflect the evolution of typical physical parameters of GRB progenitors.

After removing the redshift dependence, the isotropic-equivalent-energy function can be reasonably fitted by a broken power-law. For the dim and bright segments, we have $\psi(E_{\rm iso})\propto E_{\rm iso}^{-0.27\pm0.01}$ and $\psi(E_{\rm iso})\propto E_{\rm iso}^{-0.87\pm0.07}$, respectively (see Fig.~3). The shape of the energy function (see Fig.~3) is similar to that of the luminosity function (see Fig.~6). Moreover, our indices of $(-0.27,~-0.87)$ are comparable to the indices of $(-0.29,~-1.02)$ reported by \citet{b9}.

The GRB formation rate as a function of redshift is also calculated. The connection between (long-duration) GRBs with broad-lined Type Ic supernovae \citep[e.g.,][]{b23,b24,Woosley06,Fan11} suggests that GRB progenitors are very massive, short-lived stars, leading to expectation that the cosmic GRB formation rate would nicely follow the cosmic star formation history \citep{b25,b26,b27,b28,b29}. So far, the observed star formation rate follows $\rho(z)\propto (1+z)^{3.4}$ for $z<1$, $\propto (1+z)^{-0.3}$ for $1<z<4$ and $\propto (1+z)^{-3.5}$ for $z>4$ \citep{b20,b21} (see Fig.~10). Our results (see Fig.~10) show that the GRB formation rate increases quickly for $z\lesssim1$, then roughly keeps constant for $1\lesssim z \lesssim4$, and finally decreases at higher redshift {\it with a power index of -3.8}  (i.e., $\propto (1+z)^{-3.8\pm 2.16}$), in good agreement with the star formation rate. In future a larger GRB sample with measured redshifts would better address this topic and make progress in the fields of massive star formation and GRB physics in the early universe.

\begin{figure}
\centering
\includegraphics[width=9cm]{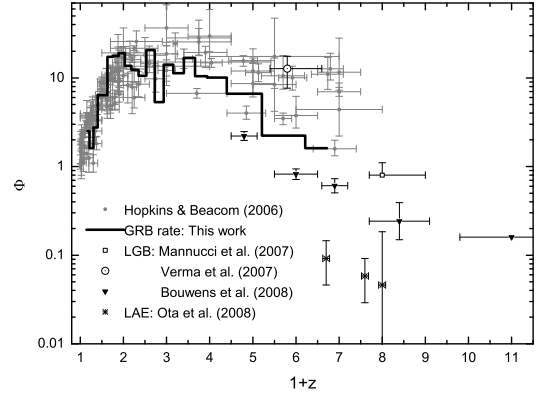}
 \caption{GRB rate comparing with SFR. The data of SFR come from \citet{b22,b20,b21}. We normalized the first point of GRB rate and SFR for convenience. \label{fig10}}
\end{figure}

Our results are based on the assumption of a power-law redshift evolution of the $E_{\rm iso}$. Besides introducing the evolution of the isotropic-equivalent-energy function, other models can also reproduce the observed data, such as one could enhance high-$z$ GRB rate by introducing a metallicity preference of GRBs or increase the total number of GRBs (e.g., Virgili et al. 2011). These models might be distinguished by the future observations.

\section*{Acknowledgments}
We are very grateful to the referee for helpful comments and suggestions. We also thank David Wanderman, Andrew Hopkins, Hansan Y\"{u}ksel and Matthew Kistler for making their data available, and Yizhong Fan and Bing Zhang for helpful discussion and suggestions. This work was supported in part by the National Natural Science Foundation of China (grants 10973041, 10921063 and 11163003). F.-W.Z. acknowledges the support by the China Postdoctoral Science Foundation funded project (No.~20110490139), the Guangxi Natural Science Foundation (No.~2010GXNSFB013050) and the doctoral research foundation of Guilin University of Technology.

\bsp

\label{lastpage}

\end{document}